# A NOVEL WINDOW FUNCTION YIELDING SUPPRESSED MAINLOBE WIDTH AND MINIMUM SIDELOBE PEAK

Md Abdus Samad

Department of Computer and Communication Engineering
International Islamic University Chittagong, Chittagong-4203, Bangladesh
asamadece@gmail.com

## ABSTRACT

*In many applications like FIR filters, FFT, signal processing and measurements, we are required (~45 dB) or less side lobes amplitudes. However, the problem is usual window based FIR filter design lies in its side lobes amplitudes that are higher than the requirement of application. We propose a window function, which has better performance like narrower main lobe width, minimum side lobe peak compared to the several commonly used windows. The proposed window has slightly larger main lobe width of the commonly used Hamming window, while featuring 6.2~22.62 dB smaller side lobe peak. The proposed window maintains its maximum side lobe peak about -58.4~-52.6 dB compared to -35.8~-38.8 dB of Hamming window for M=10~14, while offering roughly equal main lobe width. Our simulated results also show significant performance upgrading of the proposed window compared to the Kaiser, Gaussian, and Lanczos windows. The proposed window also shows better performance than Dolph-Chebyshev window. Finally, the example of designed low pass FIR filter confirms the efficiency of the proposed window.*

## 1. INTRODUCTION

The ideal approach to the design of discrete-time infinite impulse responses (IIR) filters involves the transformation of a continuous-time filter into a discrete-time filters meeting some prescribed specifications. This is partly because continuous –time filter design was highly advanced art before discrete-time filters were of interest [1]. In [2] different windows has been given with their classification such as fixed, adjustable window functions, weight windows based on Atomic functions, polynomial windows and, Z-window functions. In [3], three different window functions Han window, Hamming window and Blackman window in a general format according to evolutionary algorithm has proposed. However, it does not have any close loop formula. For window method, frequency response of a filter obtained by a periodic continuous convolution of the ideal filter in frequency domain with the Fourier transform of the window. The most straightforward approach to obtain a causal finite impulse response (FIR) is to truncate the ideal response. If $h_d[n]$ is the impulse response of the desired (ideal) IIR system, the simplest way to obtain a casual FIR filter is to define a new system with impulse response $h[n]$ of length $M+1$ as:

$$h[n] = h_d[n]w[n] \qquad (1)$$

$$w[n] = \begin{cases} f(n) & 0 \le n \le M \\ 0 & Otherwise \end{cases} \qquad (2)$$

Where *f(n)* is a function of *n* and for different function, *w[n]* have different characteristics; some of them are: Bartlett, Hanning, Hamming, Blackman, Lanczos, Kaiser, Gaussian, and Dolph-Chebyshev windows [4]-[9]. In other words, we can represent *h[n]* as the product of the desired response $h_d[n]$ and a finite-duration "window", *w[n]*. Therefore, the Fourier transform (FT) of *h[n]*, $H(e^{j\omega})$ is the periodic convolution of the desired frequency response, with FT of the window, $W(e^{j\omega})$. As a result, the FT of *h[n]* will be smeared version of the FT of $h_d[n]$. In the application, it is desired for a window function to have characteristics of smaller ripple ratio and narrower main lobe width. However, these two requirements are contradictory [1]. For the equal length of *M*+1, Hamming window offers the smallest peak of side lobe as well as main lobe width compared to Bartlett and Hanning window. The Blackman window has wider main lobe width but smaller side lobe peak compared to the Hamming window. Lanczos window [10], shows different characteristics in the main lobe depending by a positive integer. The Kaiser and Gaussian windows are tunable functions, and there is a trade-off between side lobe peaks and main lobe widths, and can be customized. The Kaiser window has the disadvantage of higher computational complexity calculating the window coefficients. Dolph-Chebyshev window has all side lobes are equal and the main lobe width is the minimum for a specific ripple ratio but it has high cost of computation. There has been great interest into the design of new windows to meet the desired specification for different applications [11]. In this paper, we present a proposed window function which has at least 6.2~22.62 *dB* less side lobe peak compared to the Hamming window, while offering smaller or equal or slightly larger main lobe width. We also show that the proposed window is better than the other windows such as Hanning, Bartlett, and Gaussian, Kaiser, Lanczos, Dolph-Chebyshev and recent proposed window in [9], [15]. We also design low pass FIR filter with the proposed window to evaluate its efficiency.

## 2. PROPOSED WINDOW FUNCTION

We have derive the proposed window by comparing is followed by comparing $\cos^L(n)$ with $\text{sinc}^L(n)$ (where *L* is integer) at least within a particular range. By comparing mathematical functions, new window was derived and proposed in [11], [12].

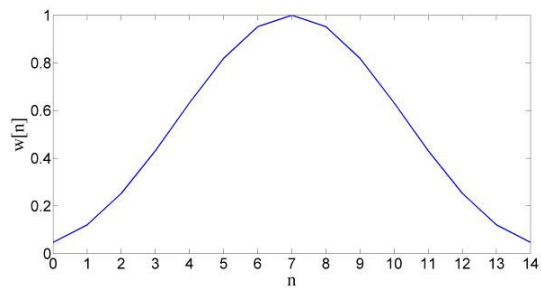

(a)

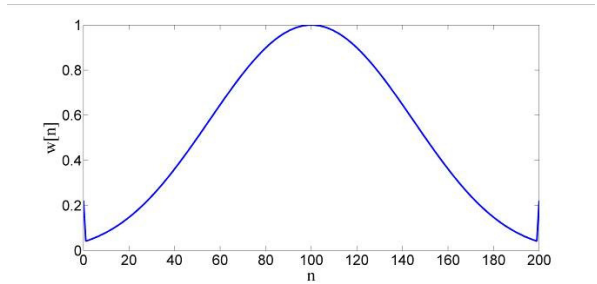

(b)

Figure 1. Shape of the proposed window for different lengths. (a) M=14, (b) M=200

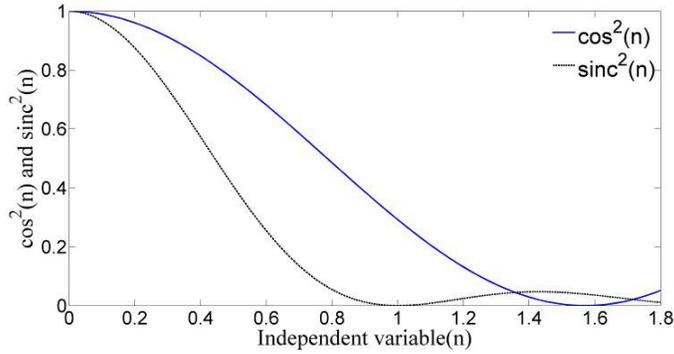

(a) $\cos^2(n)$ and $\mathrm{sinc}^2(n)$ vs $n$

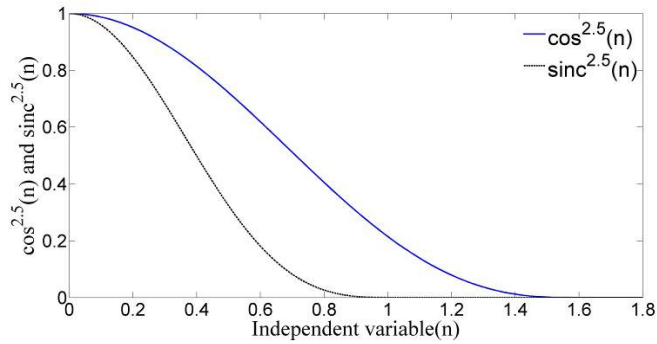

(b) $\cos^{2.5}(n)$ and $\mathrm{sinc}^{2.5}(n)$ vs $n$

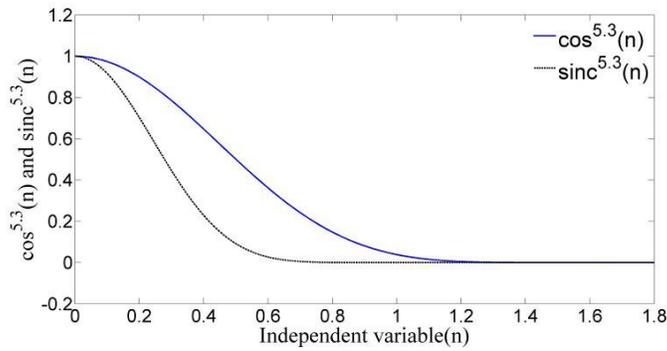

(c) $\cos^{5.3}(n)$ and $\mathrm{sinc}^{5.3}(n)$ vs $n$

Figure 2. (a), (b), and (c) comparison of $\sin^L c(n)$ and $\cos^L(n)$ vs $n (\in N)$.

The same procedure has been followed here to derive a new window function. The optimum value of window function for FIR filter design was calculated in [9] and a new window was proposed based on their findings in terms of sinc function. The optimized values of windows shows abrupt behaviour for some of these components when M is large (Figure 1.b.) around n=0 and M and for lower M the optimized components do not shown (Figure1.a) this abrupt behavior [9]. Figure 2 shows the similarity between $\sin^L c(n)$ and $\cos^L (n)$. Therefore, based on this similarity, we have examined the Lanczos window function ($\sin^L c(n)$ type) by $\cos^L (n)$ function. In addition, an important behavior from these Figure's is that as the values of L increases the two functions presents become more similar property. Based on this of similarity a new window proposed in terms $\cos^L (n)$ where the optimum values of window function has generated according to [9]. Finally, suitable formula has derived. It was observed that, the following window function fit with the optimized components for *M<19*.

$$w[n] = \begin{cases} \cos^{5.3}(2n/M - 1) & 0 \le n \le M \\ 0 & otherwise \end{cases} \tag{3}$$

But if M is large ($M \ge 20$) then,

$$w[n] = \begin{cases} 4.6051E - 9M^3 + 1.8899E - 6M^2 \\ + 0.007339M + 0.036034 & n = 0, M \\ \cos^{5.3}(2n/M - 1) & 0 < n < M \\ 0 & otherwise \end{cases} \tag{4}$$

For *n=0, M* a polynomial was proposed to fit the nonlinearity as shown in Figure 1.b. The new window yields narrower main lobe width and minimum side lobe peak comparative to other window.

The new window has the property that:

$$w[n] - \begin{cases} w[M - n] & 0 < n < M \\ 0 & otherwise \end{cases} \tag{5}$$

i.e. it is symmetric about the point M/2 and consequently has a generalized linear phase. Fig's. 1. shows the shape of the proposed window for different values of M. For $M \le 34$, w[0]=w[M] are smaller than w[1]=w[M-1], respectively and the window is like a bell-shaped function. With $M \ge 34$, w[0] and w[M] are larger than w[1] and w[M-1], respectively.

## 3. COMPARISON WITH OTHER WINDOWS

In this section, we compare the performance of the proposed window with several commonly used windows with MATLAB [13], [14].

### 3.1. Hamming Window

It has the shape of:

$$w_H = \begin{cases} 0.54 - 0.46\cos(2\pi n/M) & n = 0 < n < M \\ 0 & otherwise \end{cases} \tag{6}$$

Figure 3.a shows that the proposed window offers *-13.76751 dB* peak of side lobe with an increased main lobe width ( ~2×.005π ) with *M=14*. With *M=50* (Figure 3.b) the proposed window has some larger main lobe width but side is much smaller. Figure 6.c demonstrates that the proposed and Hamming window has approximately equal main lobe width but the proposed window offers -48.811dB-(-42.357dB) =-6.454dB less side lobe peak. The reduction in side lobe peak is 23.08151dB with *M*=10 (-58.96933*dB* compared to *-35.88782 dB*). It shows that for $10 \le n \le 200$, the side lobe peak of the proposed window compared to that of the Hamming window is -23.081512dB~6.2dB smaller. Therefore, the proposed window offers slight larger main lobe to that of the Hamming window while offering much less side lobe peak. It also reveals that, in the case of side lobe peak, the proposed window is also better than Bartlett and Hanning windows.

Table 1: Frequency response domain comparison of the proposed and Hamming window

|  | Proposed Window | | Hamming Window | |
|---|---|---|---|---|
|  | Main lobe width(-3dB) | Normalized Side lobe peak (dB) | Main lobe width(-3dB) | Normalized Side lobe peak (dB) |
| M=10 | 2×0.28906π | -58.44563 | 2×0.27344π | -35.82400 |
| M=14 | 2×0.20313π | -52.62333 | 2×0.1875π | -38.85582 |
| M=50 | 2×0.050781π | -48.46017 | 2×0.05078π | -42.47663 |
| M=50 | 2×0.050781π | -48.46017 | 2×0.050781π (M=47) | -42.23333 (M=47) |

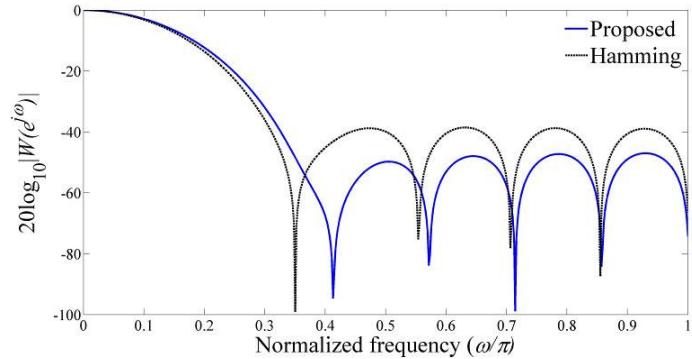

(a) Proposed M=14, Hamming M=14

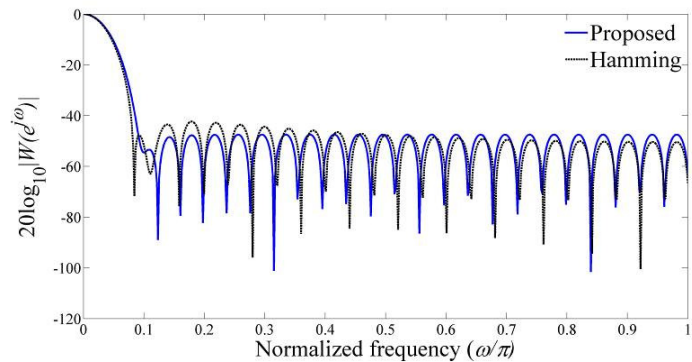

(b) Proposed M=50, Hamming M=50

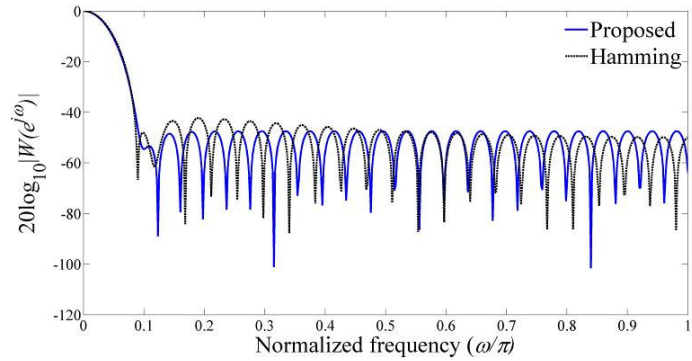

(c) Proposed M=50, Hamming M=47

Figure 3. Fourier transforms of the proposed and Hamming windows for different lengths.

## 3.2. Kaiser Window

The Kaiser window has the following shape:

$$
w_K = \begin{cases} \dfrac{I_0[\beta(1-[(n-M/2)/(M/2)^2]^{2.5}]}{I_0(\beta)} & 0 < n < M \\ 0 & otherwise \end{cases} \tag{7}
$$

where $\beta$ is the tuning parameter of the window to it shows a trade-off between the desired "side lobe peak- main lobe width," and $I_0(.)$ is the zero order modified Bessel function of the first kind. From simulated result, it is observed that for $M=50$ and $\beta=6.55$, the two windows have same side lobe peak ($\sim -48.1\ dB$) while the proposed window gives less main lobe width.

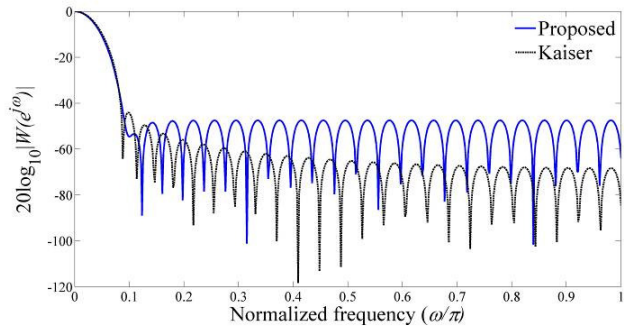

Proposed M=50, Kaiser ($\beta$=6) M=50

Figure 4. Fourier transforms of proposed window and Kaiser window with $M=50$.

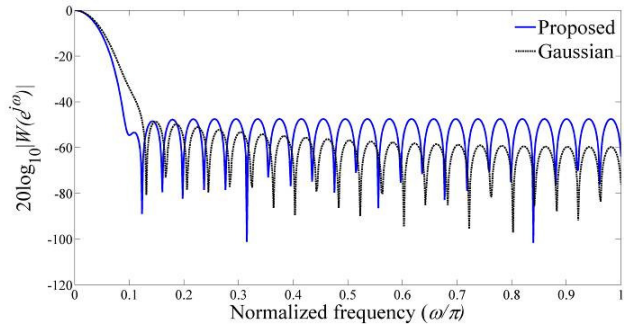

(a) Proposed M=50, Gaussian (σ=0.373) M=50

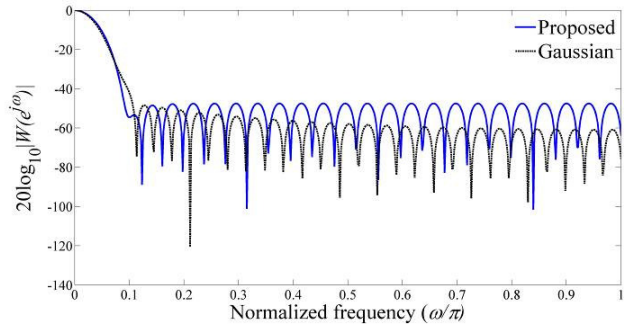

(b) Proposed M=50, Gaussian (σ=0.373) M=58

Figure 5. (a), (b) Fourier transforms of the proposed window and Gaussian (*σ=0.373*) windows with *M*=50.

If we want the Kaiser window to have the same main lobe of the proposed window, then its length should be M+1=53. Therefore, the proposed window offers the desired specifications with lower length. By decreasing *β* to *6* it also shows less side lobe peak -48.50949dB compared to -44.22365dB (-4.28584dB less), while maintaining the same main lobe width and *M* as shown in Figure 4.

### 3.3. Gaussian Window

The Gaussian window is of the form

$$
w_G = \begin{cases} e^{-\frac{1}{2}(\frac{n-M/2}{\sigma M/2})^2} & 0 \le n \le M \\ 0 & otherwise \end{cases}
\tag{8}
$$

where *σ<0.5* is the tuning parameter of the window to have the desired "main lobe width – side lobe peak" trade-off.

By setting *σ=0.373* for this window, Figure 5.(a)-(b) depicts that the side lobe peak of the two windows is *-48.85843 dB* (*M=50*), while the main lobe width of the proposed window is much less than that of the Gaussian window. Our analysis show that, for the approximately equal main lobe width, the Gaussian window need extra *8* point i.e. *M=58*.

### 3.4. Dolph-Chebyshev Window

This window can be expressed as a cosine series [5]:

$$w[n] = \frac{1}{M+1}[T_M(x_0) + \sum_{i=1}^{M/2} T_M(x_0 \cos \frac{i\pi}{M+1}) \cos \frac{2i\pi}{M+1}] \qquad (9)$$

where $T_M(x)$ is the Chebyshev polynomial of degree $M$ and $x_0$ is a function of side lobe peak and $M$. Dolph-Chebyshev window has high cost of computation, but it's important property is that all side lobes are equal and the main lobe width is minimum that can be achieved for a given ripple ratio. Figure 6, we observe that, the proposed window has a little better performance than Dolph-Chebyshev window (*0.5 dB* higher side lobe peak), but with a greater main lobe width of $\sim 2 \times 0.01\pi$. However, note that the proposed window coefficients can be computed easier than Dolph-Chebyshev window.

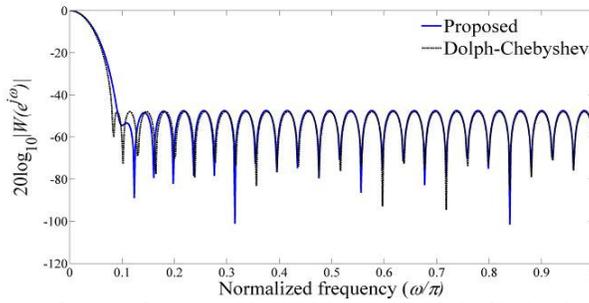

Figure 6. Fourier transforms of proposed window and Dolph-Chebyshev window with *M*=50.

### 3.5. Window proposed in [9]

The proposed window [9] is of the form:

$$w[n] = \begin{cases} 0.02 + 0.001M + (2M + 50)^{-1} & n = 0, M \\ \sin c^{2.5}(\dfrac{n - M/2}{0.654M}) & 0 < n < M \\ 0 & otherwise \end{cases} \qquad (10)$$

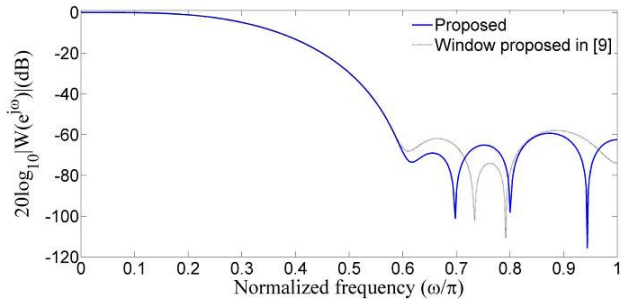

Proposed M=14, window proposed in [9] M=14

Figure 7. Fourier transforms of the proposed window (*M=14*) and window proposed in [9]. With *M=14* the proposed window offers about *-2.7 dB* peak side lobe (Figure 7) reduction than the proposed window in [9].

### 3.6. Lanczos Window

The Lanczos window has the form of:

$$w_L[n] = \begin{cases} \sin c^L (2n/M - 1) & 0 \le n \le M \\ 0 & otherwise \end{cases} \tag{11}$$

where $L$ is a positive integer number. With $M$=50, Figs 8.a and 8.b compare the proposed and Lanczos windows for $L$=1 and 2, respectively.

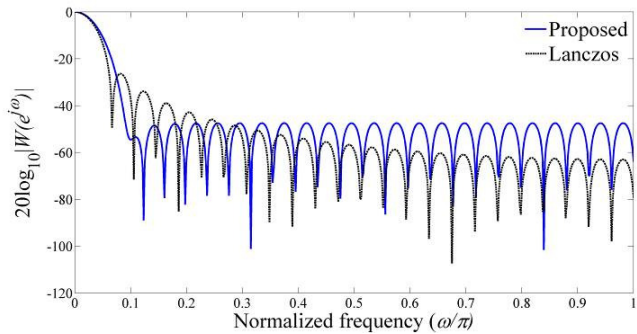

(a) Proposed M=50, Lanczos (L=1) M=50

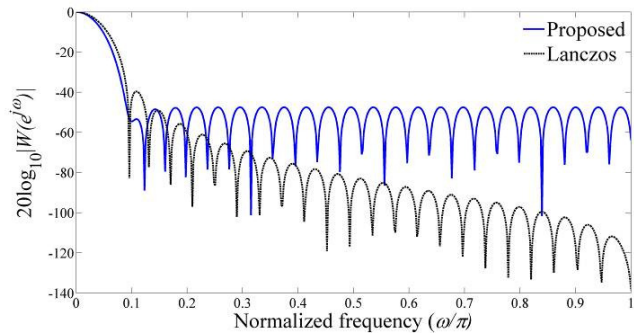

(b) Proposed M=50, Lanczos (L=2) M=50

Figure 8. Fourier transforms of proposed window and Lanczos window with *M=50*.

Figure 8.a shows that the side lobe peak of the proposed Lanczos window is *-21.1 dB* less than that of the Lanczos window, while its main lobe is just a little wider $2\times0.03\pi$ ; therefore for *L*=2, the proposed window has both less main lobe width and side lobe peaks. Figure 8.b, demonstrates that main lobe width increased with $2\times0.004\pi$ in case of Lanczos window compared to proposed window and increased side lobe peak by *-7.9 dB*.

### 3.7. Window proposed in [15]

The form of the proposed window is:

$$w[n] = 0.536 - 0.461\cos(2\pi n / M) - 0.003\cos(6\pi n / M) \quad 0 \le n \le M \tag{12}$$

For M=50 the Fourier transforms of the proposed window shows (Figure 9) about equal Mainlobe width and but reduced Sidelobe peak than the window by the equation (12).

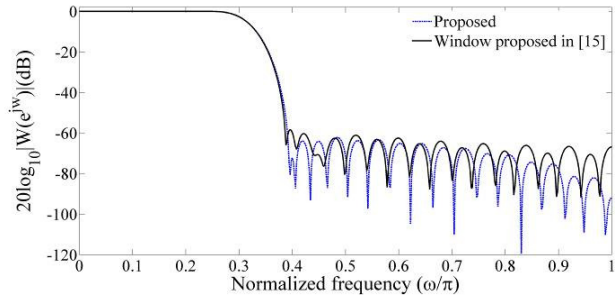

Proposed M=50, Window proposed in [15] M=50

Figure 9. Fourier transforms of the proposed window and window proposed in [15].

## 4. PERFORMANCE ANALYSIS

To study the efficiency of the proposed window we have compared the results by observing the Fourier Transform of a low pass FIR filter designed by truncating of an ideal IIR low pass filter. Having a cut off frequency of $\omega_C$, the impulse response of an ideal low pass filter is:

$$h_{LPF,ideal}[n] = \sin(\pi n) / \pi n \tag{13}$$

By windowing this IIR filter with the windows discussed in this paper, different FIR filters can be obtained. With cut off frequency $\omega_C=0.2\pi$, Figs. 10.a-10.g show the frequency response of the FIR filters designed by applying different windows of length M+1=51. Figure 10.a demonstrates that the filter achieved by the proposed window has 17.44 dB less side lobe peak than the Hamming window (M=50). This value is about 2.2 dB for Kaiser window function. This side lobe reduction is almost zero for Gaussian window but it shows larger main lobe width as shown in Figure 10.c. Figure 10.d-10.e demonstrate that the side lobe of the proposed window is smaller but main lobe is larger than Dolph-Chebyshev and  window proposed in [9]. However, in Figure 10.g the proposed window length is 50 but the window proposed in [15] is for M=47 and the frequency shows that the Sidelobe of the proposed window is more attenuated than the window proposed in [15].

## 5. CONCLUSION

The proposed window is symmetric and shows better equiripple property. Performance comparison of the proposed window compared to that of the Hamming and Kaiser window shows that the proposed window offers less side lobe with the same main lobe width. This value is almost zero for Gaussian window but the proposed window offers suppressed main lobe width than Gaussian window. It is obvious that the proposed window also outperforms the Hanning and Bartlett windows because Hamming window shows better side lobe reduction than these windows for the same main lobe width.

Table 2: Side lobe peak attenuations (dB) of the FIR filters $\omega_c = 0.2\pi$ obtained by windowing of an ideal low pass filter with different window lengths.

|  | M=50 | M=70 | M=100 | M=200 |
|---|---|---|---|---|
| Hamming | -51.37 | -52.09 | -52.69 | -53.55 |
| Kaiser ($\beta$=6) | -66.61 | -61.95 | -62.22 | -62.75 |
| Gaussian | -66.70 | -66.09 | -65.63 | -65.36 |
| Dolf-Chebyshev | -61.52 | -61.39 | -61.59 | -63.88 |
| window proposed in [9] | -60.84 | -61.37 | -62.26 | -62.11 |
| window proposed in [15] | -58.70 | -59.10 | -60.80 | -60.68 |
| **Proposed** | **-68.81** | **-67.97** | **-67.28** | **-66.56** |

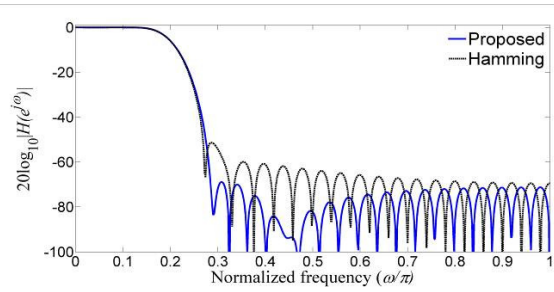

(a) Proposed M=50, Hamming M=50

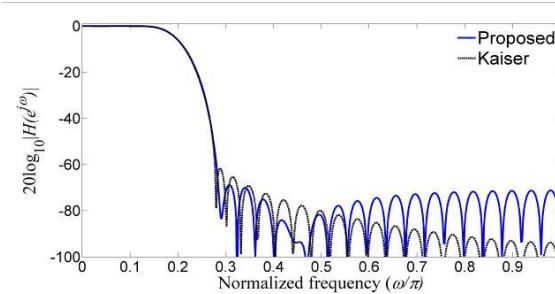

(b) Proposed M=50, Kaiser ($\beta$=6) M=50

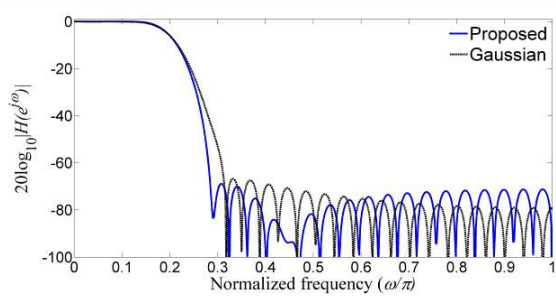

(c) Proposed M=50, Gaussian ($\sigma$=0.373) M=50

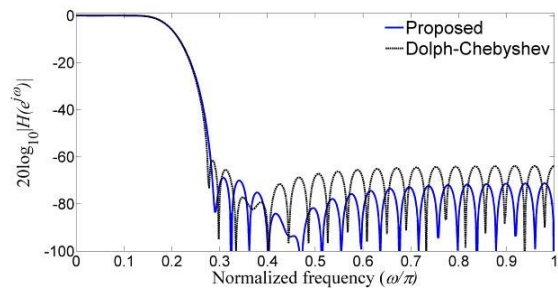

(d) Proposed M=50, Gaussian (attenuation -48dB) M=50

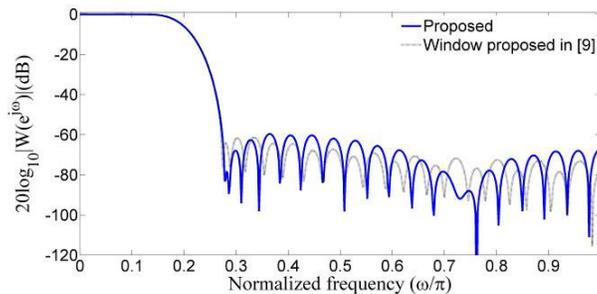

(e) Proposed M=50, window proposed in [9] M=50

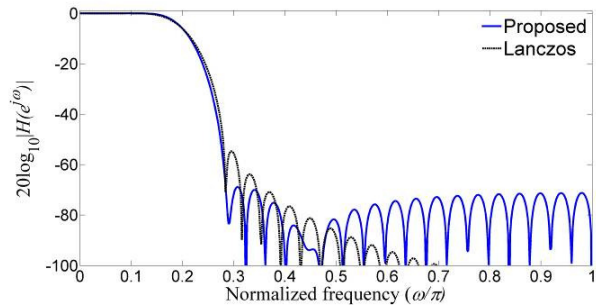

(f) Proposed M=50, Lanczos (L=2) M=50

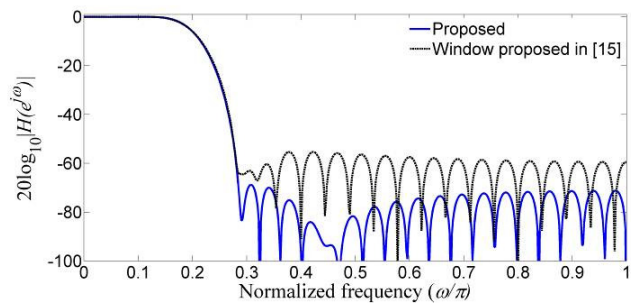

(g) Proposed M=50, window proposed in [15] M=47

Figure 10. Performance of low pass FIR filter, *M*=50, obtained by windowing of an IIR filter with different windows with cut of frequency $\omega_C = 0.2\pi$.

The designed low pass FIR filter using the proposed window achieves less ripple ratio compared to the above-mentioned window filters. Finally, for the same specification the proposed window shows more side lobe reduction with slight increased main lobe width in comparison with window proposed in [9] and with $M<19$ it shows better side lobe reduction and main lobe width comparative to other these windows for M>20 .

## Author

**Md Abdus Samad** received his B.Sc. degree in Electronics and Communication Engineering from Khulna University, Khulna, Bangladesh, in 2007. He is working as lecturer of the Faculty of Science and Engineering, International Islamic University Chittagong, Bangladesh since August 10, 2008. His research interest includes digital signal processing, wireless networks, optimization of wireless networks.

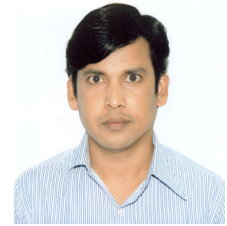